\documentclass[12pt,letterpaper]{article}
\usepackage{times}
\usepackage[margin=1in]{geometry}
\usepackage{color}
\usepackage{comment}
\usepackage{graphicx}
\usepackage[utf8]{inputenc}
\usepackage{aas_macros}
\usepackage[utf8]{inputenc}
\usepackage{textcomp}
\usepackage{enumitem,amssymb}
\usepackage{ragged2e}
\newlist{thematic}{itemize}{8}
\setlist[thematic]{label=$\square$}
\usepackage{pifont}

\usepackage{subfigure}
\usepackage{xcolor}
\usepackage{soul}
\usepackage{comment}
\usepackage{graphicx}
\usepackage{floatrow}
\usepackage{amsmath}
\usepackage[T1]{fontenc} 
\usepackage{ amssymb }
\newcommand{\ANLHEP}{HEP Division, Argonne National Laboratory, Lemont, IL 60439, USA}

\newcommand{\BenGurion}{Department of Physics, Ben-Gurion University, Be'er Sheva 84105, Israel}
\newcommand{\Bicocca}{University of Milano - Bicocca, Piazza della Scienza, 3, Milano, Italy}
\newcommand{\Birm}{School of Physics and Astronomy and Institute of Gravitational Wave Astronomy, University of Birmingham, Edgbaston B15 2TT, UK}
\newcommand{\Bohr}{The Niels Bohr Institute \& Discovery Center, Blegdamsvej 17, DK-2100 Copenhagen, Denmark}
\newcommand{\BNL}{Brookhaven National Laboratory, Upton, NY 11973}
\newcommand{\Brown}{Brown University, Providence, RI 02912}

\newcommand{\BU}{Boston University, Boston, MA 02215}

\newcommand{\Caltech}{California Institute of Technology, Pasadena, CA 91125}

\newcommand{\CPPM}{Aix Marseille Univ, CNRS/IN2P3, CPPM, Marseille, France}

\newcommand{\CIERA}{Center for Interdisciplinary Exploration and Research in Astrophysics (CIERA) and Department of Physics and Astronomy, Northwestern University, Evanston, IL 60208}
\newcommand{\CfA}{Harvard-Smithsonian Center for Astrophysics, MA 02138}

\newcommand{\CITA}{Canadian Institute for Theoretical Astrophysics, University of Toronto, Toronto, ON M5S 3H8, Canada}

\newcommand{\CMUCosmo}{Department 
of Physics, McWilliams Center for Cosmology, Carnegie Mellon University}

\newcommand{\EPFL}{Institute of Physics, Laboratory of Astrophysics, Ecole Polytechnique Fédérale de Lausanne (EPFL), Observatoire de Sauverny, 1290 Versoix, Switzerland}

\newcommand{\FNAL}{Fermi National Accelerator Laboratory, Batavia, IL 60510}

\newcommand{\GRAPPA}{GRAPPA Institute, University of Amsterdam, Science Park 904, 1098 XH Amsterdam, The Netherlands}

\newcommand{\IAP}{Institut d'Astrophysique de Paris (IAP), CNRS \& Sorbonne University, Paris, France}

\newcommand{\IBS}{Institute for Basic Science (IBS), Daejeon 34051, Korea}

\newcommand{\ICE}{Institute of Space Sciences (ICE, CSIC), Campus UAB, Carrer de Can Magrans, s/n, 08193 Barcelona, Spain}

\newcommand{\IFT}{Instituto de Fisica Teorica UAM/CSIC, Universidad Autonoma de Madrid, 28049 Madrid, Spain}

\newcommand{\INFNFE}{Istituto Nazionale di Fisica Nucleare, Sezione di Ferrara, 44122 Ferrara, Italy}

\newcommand{\JPL}{Jet Propulsion Laboratory, California Institute of Technology, Pasadena, CA, USA}
\newcommand{\KASSI}{Korea Astronomy and Space Science Institute, Daejeon 34055, Korea}

\newcommand{\KICP}{Kavli Institute for Cosmological Physics, Chicago, IL 60637}
\newcommand{\KIPAC}{Kavli Institute for Particle Astrophysics and Cosmology, Stanford 94305}

\newcommand{\LBL}{Lawrence Berkeley National Laboratory, Berkeley, CA 94720}

\newcommand{\LLNL}{Lawrence Livermore National Laboratory, Livermore, CA, 94550}

\newcommand{\MIT}{Massachusetts Institute of Technology, Cambridge, MA 02139}

\newcommand{\LUPM}{Laboratoire Univers et Particules de Montpellier, Univ. Montpellier and CNRS, 34090 Montpellier, France}
\newcommand{\NAOC}{National Astronomical Observatories, Chinese Academy of Sciences, PR China}

\newcommand{\Nottingham}{University of Nottingham, NG7 2RD Nottingham, United Kingdom}

\newcommand{\ParisSud}{Universit\'{e} Paris-Sud, LAL, UMR 8607, F-91898 Orsay Cedex, France \& CNRS/IN2P3, F-91405 Orsay, France}
\newcommand{\PI}{Perimeter Institute, Waterloo, Ontario N2L 2Y5, Canada}

\newcommand{\Port}{Institute of Cosmology \& Gravitation, University of Portsmouth, Dennis Sciama Building, Burnaby Road, Portsmouth PO1 3FX, UK}
\newcommand{\Princeton}{Princeton University, Princeton, NJ 08544}

\newcommand{\Queensland}{The University of Queensland, School of Mathematics and Physics, QLD 4072, Australia}

\newcommand{\Rice}{Department of Physics \& Astronomy, Rice University, Houston, Texas 77005, USA}

\newcommand{\Sejong}{Department of Physics and Astronomy, Sejong University, Seoul, 143-747, Korea}

\newcommand{\Siena}{Siena College, 515 Loudon Road, Loudonville, NY 12211, USA}
\newcommand{\SimonFraser}{Department of Physics, Simon Fraser University, Burnaby, BC V5A 1S6, Canada}

\newcommand{\SMU}{Southern Methodist University, Dallas, TX 75275}

\newcommand{\SoCal}{Department of Physics and Astronomy, University of Southern California, Los Angeles, CA 90089, USA}
\newcommand{\Stanford}{Stanford University, Stanford, CA 94305}

\newcommand{\SussexAstronomy}{Astronomy Centre, School of Mathematical and Physical Sciences, University of Sussex, Brighton BN1 9QH, United Kingdom}

\newcommand{\Tamu}{Texas A\&M University, College Station, TX 77843 }

\newcommand{\UChicago}{University of Chicago, Chicago, IL 60637}
\newcommand{\UCI}{University of California, Irvine, CA 92697}
\newcommand{\UCLA}{University of California at Los Angeles, Los Angeles,  CA 90095}
\newcommand{\UCL}{University College London, WC1E 6BT London, United Kingdom}

\newcommand{\UGTO}{Divisi\'on de Ciencias e Ingenier\'ias, Universidad de Guanajuato, Le\'on 37150, M\'exico}

\newcommand{\UMich}{University of Michigan, Ann Arbor, MI 48109}

\newcommand{\UoM}{Jodrell Bank Center for Astrophysics, School of Physics and Astronomy, University of Manchester, Oxford Road, Manchester, M13 9PL, UK}
\newcommand{\UPenn}{Department of Physics and Astronomy, University of Pennsylvania, Philadelphia, Pennsylvania 19104, USA}
\newcommand{\UR}{Department of Physics and Astronomy, University of Rochester, 500 Joseph C. Wilson Boulevard, Rochester, NY 14627, USA}

\newcommand{\UWaterloo}{Department of Physics and Astronomy, University of Waterloo, 200 University Ave W, Waterloo, ON N2L 3G1, Canada}

\newcommand{\WCA}{Centre for Astrophysics, University of Waterloo, Waterloo, Ontario N2L 3G1, Canada}

\newcommand{\WVU}{CSEE, West Virginia University, Morgantown, WV 26505, USA}
\newcommand{\WVUGWAC}{Center for Gravitational Waves and Cosmology, West Virginia University, Morgantown, WV 26505, USA}

\newcommand{\YorkU}{Department of Physics and Astronomy, York University, Toronto, ON M3J 1P3, Canada}

\usepackage[maxbibnames=3,
            natbib=true,
            sorting=none,backend=bibtex]{biblatex}
\begin{document}

\LARGE
{\center\noindent Astro2020 Science White Paper \\[0.5em]
\noindent Gravitational wave cosmology and astrophysics with large spectroscopic galaxy surveys\\}

\normalsize

\vspace{0.4cm}

\noindent \textbf{Thematic Areas:} Cosmology and Fundamental Physics, Formation and Evolution of Compact Objects, Multi-Messenger Astronomy and Astrophysics \\

\noindent \textbf{Principal Author:}\\
Name: Antonella Palmese     \\
Institution:  Fermi National Accelerator Laboratory\\
Email: \url{palmese@fnal.gov}\\

\noindent \textbf{Co-authors:} Or Graur$^{1}$, 
James T. Annis$^{2}$, 
Segev BenZvi$^{3}$, 
Eleonora Di Valentino$^{4}$, 
Juan Garcia-Bellido$^{5}$, 
Satya {Gontcho A Gontcho}$^{3}$, 
Ryan Keeley$^{6}$, 
Alex Kim$^{7}$, 
Ofer Lahav$^{8}$, 
Samaya Nissanke$^{9}$, 
Kerry Paterson$^{10}$, 
Masao Sako$^{11}$, 
Arman Shafieloo$^{6}$, 
Yu-Dai Tsai$^{2}$\\

\noindent \textbf{Endorsers:}\\ Mustafa A. Amin$^{12}$, 
Robert Armstrong$^{13}$, 
Jacobo Asorey$^{6}$, 
Arturo Avelino$^{1}$, 
Kevin Bandura$^{14,15}$, 
Elizabeth Buckley-Geer$^{2}$, 
Francisco J Castander$^{16}$, 
Christopher J. Conselice$^{17}$, 
Asantha Cooray$^{18}$, 
Matteo Cremonesi$^{2}$, 
Rupert A. C. Croft$^{19}$, 
Tamara M Davis$^{20}$, 
Kelly A. Douglass$^{3}$, 
Duan Yutong$^{21}$, 
Stephanie Escoffier$^{22}$, 
Giulio Fabbian$^{23}$, 
Arya Farahi$^{19}$, 
Wen-fai Fong$^{10}$, 
Martina Gerbino$^{24}$, 
William Hartley$^{8}$, 
Adam J. Hawken$^{22}$, 
Lars Hernquist$^{1}$, 
Dragan Huterer$^{25}$, 
Johann Cohen-Tanugi$^{26}$, 
Kenji Kadota$^{27}$, 
Robert Kehoe$^{28}$, 
Jean-Paul Kneib$^{29}$, 
Savvas M. Koushiappas$^{30}$, 
Ely D.~Kovetz$^{31}$, 
Benjamin L'Huillier$^{6}$, 
Massimiliano Lattanzi$^{32}$, 
Pablo Lemos$^{8}$, 
Andr\'es A. Plazas$^{33}$, 
Raffaella Margutti$^{10}$, 
Jennifer L. Marshall$^{34}$, 
Kiyoshi Masui$^{35}$, 
James Mertens$^{36,37,38}$, 
John Moustakas$^{39}$, 
Suvodip Mukherjee$^{40}$, 
Pavel Naselsky$^{41}$, 
Federico Nati$^{42}$, 
Gustavo Niz$^{43}$, 
Andrei Nomerotski$^{44}$, 
Lyman Page$^{33}$, 
Will~J. Percival$^{45,46,37}$, 
Elena Pierpaoli$^{47}$, 
Levon Pogosian$^{48}$, 
Giuseppe Puglisi$^{49,50}$, 
Marco Raveri$^{51,52}$, 
Gra\c{c}a Rocha$^{53,54}$, 
Graziano Rossi$^{55}$, 
Alberto Sesana$^{56}$, 
Sara Simon$^{25}$, 
Aritoki Suzuki$^{7}$, 
Matthieu Tristram$^{57}$, 
Nathan Whitehorn$^{58}$, 
Zhilei Xu$^{11}$, 
Gong-Bo Zhao$^{59,60}$, 
Ningfeng Zhu$^{11}$\\
{\small \emph{(Affiliations are listed at the end of the paper)}}\\

\noindent \textbf{Abstract:} During the next decade, gravitational waves will be observed from hundreds of binary inspiral events. When the redshifts of the host galaxies are known, these events can be used as ``standard sirens'', sensitive to the expansion rate of the Universe. Measurements of the Hubble constant $H_0$ from standard sirens can be done independently of other cosmological probes, and events occurring at $z<0.1$ will allow us to infer $H_0$ independentently of cosmological models. The next generation of spectroscopic galaxy surveys will play a crucial role in reducing systematic uncertainties in $H_0$ from standard sirens, particularly for the numerous ``dark sirens'' which do not have an electromagnetic counterpart. In combination with large spectroscopic data sets, standard sirens with an EM counterpart are expected to constrain $H_0$ to $\sim 1-2\%$ precision within the next decade. This is competitive with the best estimates of $H_0$ obtained to date and will help illuminate the current tension between existing measurements.

\newpage

\section{Introduction}\label{sec:sciencecases}

The discovery of the gravitational wave (GW) signal GW170817 by LIGO/Virgo and its electromagnetic (EM) counterpart \cite{MMApaper} has opened a new era of GW cosmology with the first measurement of the Hubble constant, $H_0$, using \textbf{standard sirens} (StS;\cite{schutz}) \cite{2017Natur.551...85A}. We review the GW science cases to which large-scale spectroscopic galaxy surveys can make valuable contributions. Throughout, we mention the Dark Energy Spectroscopic Instrument (DESI; \cite{2016arXiv161100036D,2016arXiv161100037D}), but it should be stressed that the science cases enumerated below are just as relevant for other existing and upcoming large scale spectroscopic galaxy surveys (e.g., Taipain, SDSS-V, and 4MOST  \cite{taipan, 2017arXiv171103234K, 2016SPIE.9908E..1OD}). 

\vspace{2mm}
``Standard candles'', such as Cepheids and Type Ia supernovae (SNe Ia), have long been used to measure $H_0$ and other cosmological parameters. StSs are another method that is independent of traditional distance-ladder approaches \cite{schutz,2005ApJ...629...15H,2010ApJ...725..496N,delpozzo,chen}. In fact, the amplitude of the GW signal depends directly on the distance $D$ to the object. The redshift $z$ of the source can be measured either from direct detection of the object and/or its host galaxy (if the GW source has an EM counterpart: \textbf{bright sirens}) or through a statistical approach using the ensemble of galaxies in the area covered by the localization uncertainty region of the GW signal (if there is no EM counterpart: \textbf{dark sirens}). Distance and redshift are related to $H_0$ through $v_H(z_H) = H_0 D$, where $v_H(z_H)$ and $z_H$ are the recession velocity and redshift of the object, respectively, due to the Universe's expansion.  For objects in the local Universe ($z \lesssim 0.1$), the relation is linear and only depends on the local value of the Hubble parameter: $cz_H= H_0 D$, where $c$ is the speed of light.

\vspace{2mm}
\textbf{\textit{A New Probe of the Hubble constant.}} Recently, two leading probes of $H_0$ have come into tension. The latest measurement of $H_0$ based on the distance ladder composed of Cepheid variables and SNe Ia is $73.48\pm1.66~{\rm km~s^{-1}~Mpc^{-1}}$ \cite{Riess_2018}. On the other hand, the latest measurement of $H_0$ from the cosmic microwave background (CMB), assuming a spatially flat $\Lambda$CDM model, is $67.27 \pm 0.60~{\rm km~s^{-1}~Mpc^{-1}}$ \cite{planck18}, $3.7\sigma$ lower.
A new, independent probe of $H_0$ could clarify this tension. Low-redshift StSs ($z \lesssim 0.1$) will help estimation of Hubble constant with very high precision during the next decade~\cite{chen}. However, using StSs at $z \gtrsim 0.1$ requires a proper knowledge of the background cosmology to avoid introducing biases in the analysis~\cite{2018arXiv181207775S}. Information from large spectroscopic surveys estimating the expansion history of the Universe in a model independent manner combined with StSs can resolve this issue.  

\vspace{2mm}
\textbf{\textit{Dark Energy.}} Despite the wealth of analyses aimed at understanding the dark sector of the Universe, the nature of dark matter and dark energy (DE) remains elusive. Beyond $z \gtrsim 0.1$, the distance--redshift relation depends on cosmological parameters beyond $H_0$, including the Universe's matter and DE density. It thus follows that GW events at those redshifts will be a new source of information to estimate the background rate of expansion. This is particularly true for the next generation of GW detectors, such as the Laser Interferometer Space Antenna (LISA, \cite{lisa}), the Einstein Telescope (ET) and the Cosmic Explorer (CE), where the increased sensitivity will allow inference of cosmological parameters from precise distance measurements of GW events to very large distances \cite{taylor,tamanini,caprini,2010CQGra..27u5006S,delpozzo18}.

\vspace{2mm}

\textbf{\textit{Formation and evolution of GW sources.}} The nature of the formation and evolution of the binaries that produce GWs is still mostly unknown. It is in fact not clear if the components are formed through a burst of star formation as a binary system (the ``isolated binary'' scenario), and if they can survive two SN explosions. In the case of black hole (BH) binaries, it is not even clear if the components observed so far can form as stellar objects, whether they are primordial BHs \cite{PhysRevLett.116.201301,2017PDU....15..142C}, the result of dynamical interaction of stellar BHs in dense environments \cite{lipunov,faber} or in quasar accretion disks. Analyses of the host galaxy of GW events can provide insights into the environments in which those system evolve, and thus can help in constraining formation and evolution scenarios of compact object binaries \cite{palmese,Belczynski}.

\section{Probes}

In this Section we further explore the probes necessary to address the science cases introduced in Section \ref{sec:sciencecases}, and present how spectroscopic surveys can contribute to such goals. We treat separately the cases in which the EM counterpart to a GW event is (or is not) identified.

\vspace{4mm}

\textbf{\underline{Bright sirens.}}
The promise of the new multi-messenger era is most prominent in GW events that are accompanied by an EM counterpart, as the host galaxy, and thus its redshift, can be identified more easily. However, peculiar velocities are one possible source of systematic bias, as they cause a deviation of measured galaxies' velocities from the Hubble flow. It is thus important to precisely measure the redshift of the event's host galaxy and of its surrounding environment in order to correctly recover peculiar velocities. In turn, these are needed to recover the velocity component, $v_H$, due solely to the Hubble flow, which enters $v_H(z_H) = H_0 D$. %

\vspace{3mm}

\textbf{\emph{Follow-up of EM counterpart candidates.}} Identification of EM counterparts is paramount to enable cosmology with bright StSs. According to recent EM counterpart searches \cite{marcelle,doctor}, programs such as the DECam GW follow-up are likely to observe $\sim 10$ interesting kilonova candidates per square degree. This increases if there are no deep galaxy catalogs available for SN rejection in that area of the sky. SNe are in fact the transients most likely to contaminate the search for kilonovae \cite{2013ApJ...767..124N}. Such contamination can be avoided by identifying candidates associated with galaxies that are too far away to be the GW host \cite{doctor}, or through color information of the transients. 

Thanks to the large field of view and number of fibers, wide-field multi-object spectroscopy is ideal to quickly follow-up interesting candidates selected by imaging surveys. This can be achieved by assigning ancillary fibers to the candidates. The expected number of EM counterpart candidates per DESI pointing, for example, should be on the order of 50--100, which is below the number of expected spare fibers ($\sim 500$).

A significant fraction of GW signals detected in the next decade may be well-localized (within $5$--$20~{\rm deg}^2$ at 90\% Confidence Level; \cite{LIGOprospects}), allowing wide-field spectroscopic instruments to cover the whole high-probability area with just a few pointings, and ideally identify the kilonova event among the counterpart candidates. 
Timely identification of kilonovae is fundamental for further follow-up (which can be pursued later with smaller field-of-view telescopes), as they are expected to fade away on a timescale of days.  Well-localized events will be a fraction of few to tens of mergers per year for the Northern or Southern sky hemisphere.

BH--NS mergers have yet to be observed, which makes the rate of this type of events uncertain. Theoretical works suggest that they may be more numerous than NS--NS mergers \cite{2018MNRAS.481.1908K}. 
The discovery of the first BH--NS merger counterpart would not only be a significant contribution to our understanding of compact object mergers, but may also allow us to place tighter constraints on $H_0$ than a NS--NS event would \cite{vitale}. In addition, if there are exotically light NS--BH coalescences (with the BH being around the NS mass) and the GW signature is degenerate with NS-NS merger, one can use these observations to infer the BH nature. This is especially crucial in searching for BHs converted from NSs by dark matter-induced collapse \cite{Bramante:2017ulk,Yang:2017gfb}.

We conclude that following up a fraction of NS--NS or BH--NS merger events (at least the well-localized ones, i.e., the ``golden events'' \cite{thone}) through assignment of spare fibers over fields that need to be observed by the survey,
would not be disruptive for the main science goals of spectroscopic galaxy surveys, but has the potential to strongly impact multi-messenger searches of the next decade. We also note that observed EM candidates that are not the kilonova of interest are likely to be SNe, and as such could provide useful measurements for SN cosmology.

\vspace{3mm}

\textbf{\emph{Host galaxies.}} Similarly to the case of transient follow-up, multi-object spectroscopy represents a unique opportunity to quickly provide a spectrum of a host galaxy and its environment. 
The $H_0$ uncertainty from one StS depends on distance and redshift uncertainties as \cite{chen,Mortlock}: 
\begin{equation}
    \sigma_H \simeq \frac{1}{D}\sqrt{c^2 \sigma_z^2+\sigma_v^2+H_0^2\sigma_D^2} \, ,\label{eq:sigmaH1}
\end{equation}
where $\sigma_z$ is the host galaxy redshift uncertainty, $\sigma_v$ is the uncertainty on the peculiar velocity, and $\sigma_D$ is the distance uncertainty from the GW data. The host galaxy redshift is the measured spectroscopic redshift (spec-$z$), which for the DESI Bright Galaxy Survey (BGS) will have a typical uncertainty of $c \sigma_z < 100 ~{\rm km~s}^{-1}$. The redshift $z_H$, due to the cosmological expansion only, has a larger uncertainty due to the difficulty in measuring galaxies' peculiar velocities at the typical distances considered for the bright events ($\lesssim 200$ Mpc). In other words, typically $\sigma_v >c\sigma_z$ for spec-$z$'s. For the highest signal-to-noise ratio (SNR) events, the distance estimate may be more precise than the other quantities in play, and the peculiar velocity error will be dominant. %
DESI and Taipan will provide redshifts for a much denser sample than previous spectroscopic surveys, allowing a more precise estimate of the peculiar velocity flow. 

\begin{figure}
    \floatbox[{\capbeside\thisfloatsetup{capbesideposition={right,top},capbesidewidth=5.2cm}}]{figure}[\FBwidth]
{\caption{\small Hubble constant uncertainty ($1\sigma$) as a function of combined GW events with associated EM counterpart. The shaded regions show the impact of the peculiar velocity uncertainty between $100$ and $400~{\rm km~s^{-1}}$ for different distance reaches $D_*$. The latest results from standard candles (SH0ES, \cite{Riess_2018}) and CMB (\emph{Planck}, \cite{planck18}) are also shown. 
    }\label{fig:sigmaH}}
    {\includegraphics[width=1.08\linewidth,trim=0.4 0 0 0,clip]{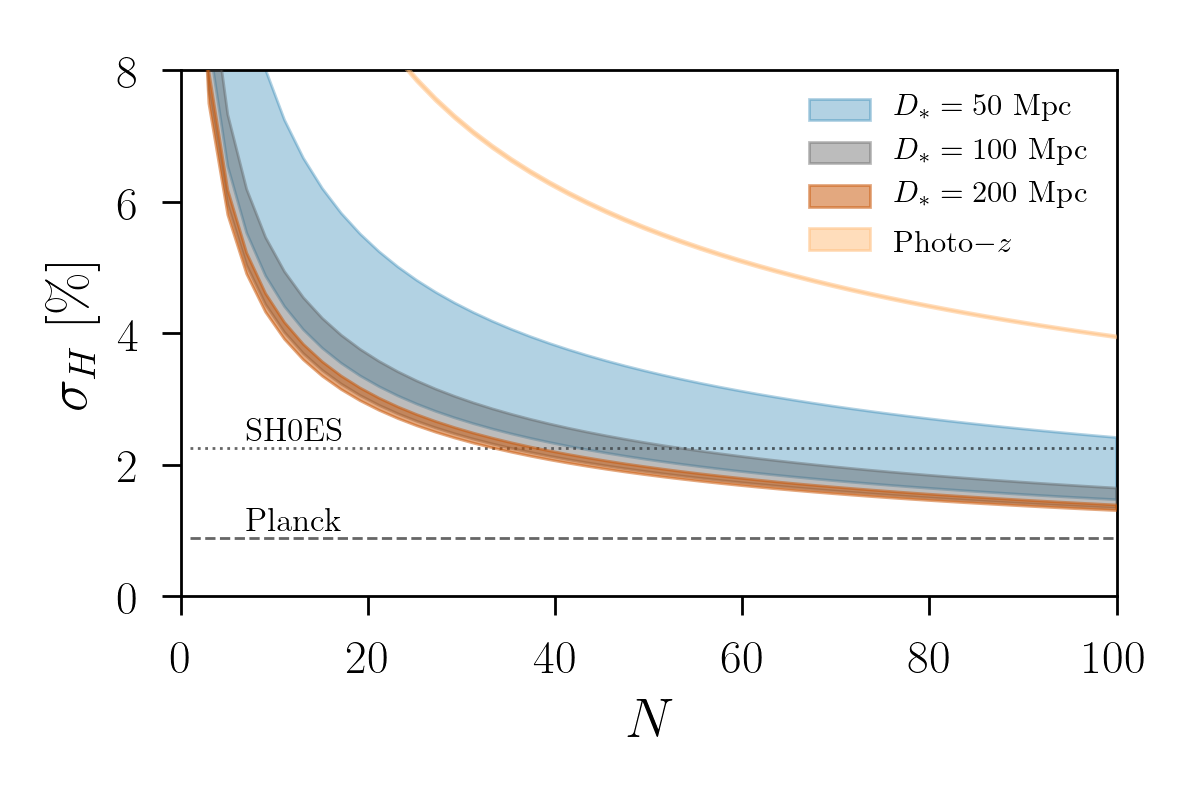}}
    \vspace{-1cm}
\end{figure}

Following \cite{Mortlock}, we show the predicted $H_0$ precision from different numbers of combined events, for different distance reaches, $D_*$,\footnote{Here $D_*$ is the distance out to which sources with the minimum SNR $\rho_*$ can be detected. We use $\rho_*=12$.} in Fig. \ref{fig:sigmaH}. The shaded regions show, for each $D_*$, the impact of the peculiar velocity uncertainty between $100$ and $400~{\rm km~s^{-1}}$. The velocity uncertainty is eventually suppressed by the increased distance reach, as most events will come from farther away where the distance uncertainty dominates. However, if we consider only the loudest GW events with a more precise distance uncertainty, or if we are only able to identify EM counterparts for the closest events, the effect of peculiar velocities will still be the most prominent. This is also true if the distance precision is improved by breaking the degeneracy between distance and inclination angle using EM data \cite{2010ApJ...725..496N}, potentially improving the $H_0$ uncertainty by a factor $2-3$ \cite{guidorzi,Hotokezaka}.

Note that, for example, at the distance of 40 Mpc (like GW170817), a bias of 200 km s$^{-1}$ in the reconstructed peculiar motion would translate into an $H_0$ bias of 5 km s$^{-1}$ Mpc$^{-1}$. A precise and accurate measurement of peculiar velocities around host galaxies is thus needed in order to perform $H_0$ analyses, and it is more important the more precise the distance measurements are. Ideally, by reaching $\sigma_v \sim 100 ~{\rm km~ s^{-1}}$, 
we can reach a precision on $H_0$ of $\sim 2\%$ regardless of the distance reach after $\sim \mathcal{O}(50)$ events. Photometric redshifts (Photo-$z$'s) will not provide competitive constraints for bright sirens, even with an ideal uncertainty of $\sigma_z\sim 0.01$ (see Fig. 1): spec-$z$'s are needed.

Analyses of host galaxy properties can also constrain the formation and evolution scenarios of NS--BH and NS--NS binaries. With this assumption, \cite{palmese} and \cite{Belczynski} used observations of the GW170817 host galaxy to suggest NS--NS formation scenarios which are alternatives to the more standard ``isolated binary'' case. \cite{Blanchard} derived a time delay constraint from the star formation history of the galaxy, assuming the NSs formed and evolved as an isolated binary. Further identification of host galaxies will be crucial in shedding light on the formation channels of these systems.

When a GW event with an EM counterpart is detected, it is possible that we will already have galaxy measurements in its localization region from DESI or other spectroscopic galaxy surveys. If that is not the case, but the event falls into a region that is planned for future observations, that region could be prioritized in near-future observations, although this type of observation does not need to be pursued in the short time scales necessary for follow-up of the EM candidate itself.  However, if a spectroscopic follow-up of EM candidates is issued, then galaxy spectra could be taken in conjunction with the candidates. Pursuing this type of science would cause minimal disruption to the planned observing strategy. 

\vspace{4mm}

\textbf{\underline{Dark sirens.}} 
The expected rate of events with an EM counterpart (NS mergers) is much lower than those that are expected to be dark, like BH mergers ($\sim 1$ to 10; \cite{GWTC}). Moreover, the EM counterpart to GW170817 was extremely bright and nearby. Generally, we expect NS merger counterparts to be more challenging to identify. Several authors have explored the possibility of making cosmological measurements without an EM counterpart (e.g., \cite{schutz,chen,fishbach,darksiren}), by taking into account a whole sample of potential host galaxies within a statistical framework. \cite{darksiren} used probable host galaxies with photo-$z$'s from the Dark Energy Survey (DES), currently the best available catalog in the BH-BH event GW170814 localization area. For a DES-like photo-$z$ precision, the statistical uncertainty on $H_0$ can reach $\sim 5\%$ after $\sim 100$ well-localized events ($\lesssim 60~{\rm deg}^2$) are combined \cite{darksiren}. Photo-$z$ systematics will likely become a dominant source of uncertainty when enough dark events can be combined to reach a statistical uncertainty on $H_0$ of $\sim 10\%$. Spectroscopic redshifts will thus be needed to improve the Hubble constant measurement. The statistical StS method is likely to become more valuable when the events localization will be improved by the addition of new interferometers to the current LIGO/Virgo network.

\vspace{4mm}

\textbf{\emph{Host galaxies.}} The DESI BGS is expected to be a premier dataset for this type of science. In fact, a complete survey of galaxies down to magnitude $r\sim 19.5$ out to $z \lesssim 0.4$, will serve as the ideal map of potential host galaxies for these events that are expected to be detectable out to $z\lesssim 0.3$ for LIGO/Virgo/KAGRA at design sensitivity \cite{LIGOprospects}. \footnote{The loudest events will actually be detected out to redshift $z\sim 1$.} 
By targeting a broad range of galaxy types, the BGS will also include those galaxies which are more likely to be binary BH hosts. According to recent studies, the hosts could be star forming, contain the most stellar mass, or have low metallicities (e.g., \cite{2014ApJ...789..120B}). 

DESI will also enable science analyses of dark events beyond derivation of the cosmological parameters. \cite{GWLSS} show how a cross-correlation of DESI galaxies with GW catalogs of binary BH mergers will be able to discern between different binary BH formation scenarios by comparing the spatial distribution of mergers versus the galaxies' distribution. The dark StS probes will not require disruption to planned observing strategies. 

\section{Conclusions \& Outlook}

Standard sirens are an extremely promising cosmological probe that can constrain the Hubble constant independently of measurements of standard candles and the CMB. 
In order to enable the science presented in this paper, we suggest the following:
 \begin{itemize}
     \item {\bf Support of GW science with large spectroscopic galaxy surveys and coordination with GW experiments.} Complete galaxy catalogs with accurate redshift measurements will be an invaluable resource in the analysis of GW events with EM counterparts, and are even more important for the statistical analyses of the much more common dark sirens. This is crucial for achieving a $\sim 2\%$($1\%$) uncertainty in $H_0$ using bright StSs, which may become possible already in the early(mid)-2020s \cite{chen}. Such a constraint will clarify the Hubble constant tension, and it will be competitive with current results from CMB and standard candles. In particular, $H_0$ estimates from bright StSs will be model-independent,\footnote{As these projections come from events at $z\lesssim0.1$.} 
     as opposed to the $\lesssim 1 \%$ precision measurements from CMB experiments, which are tied to a flat $\Lambda$CDM scenario \cite{divalentino}. Dark StSs will be an alternative when the EM counterpart cannot be identified. These type of events are expected to provide a less stringent ($\lesssim10\%$) precision \cite{chen,darksiren} in the next decade, but have the potential of probing dark energy with next generation GW experiments  \cite{2011ApJ...732...82P}. The same galaxy catalogs (e.g., the DESI main survey will observe large red galaxies out to $z\sim 1$) will also be useful in the 2030s for host identification of GW events from LISA, ET, and CE, which will reach much larger distances. Large spectroscopic galaxy catalogs from experiments planned for the 2020s will also act as important pathfinder to cross-correlate with GW catalogs of BH mergers \cite{GWLSS}. The galaxy surveys in question may have already observed (or planned to target) the galaxies of interest, thus this goal can be met with minimal disruption to the survey strategy. 
\item {\bf Dedicate a fraction of observing time to follow-up well-localized GW counterpart candidates.} An improved statistics of StSs is in fact the first challenge to achieve the $H_0$ precision asserted above (see Figure 1). %
It is possible that the EM counterpart identification will not be feasible on short timescales without wide-field multi-object spectroscopy. Moreover, measurements of the EM counterpart can improve the $H_0$ constraints presented above by a factor of 2--3 by breaking the degeneracy between distance and inclination angle \cite{Hotokezaka}. Beyond cosmology, the overlap between GW and EM observations promises exciting new discoveries, including the first BH--NS EM counterpart and an improved understanding of the physics of compact object binaries. 

 \end{itemize}

\printbibliography
\section*{Affiliations}
$^{1}$ \CfA \\$^{2}$ \FNAL \\$^{3}$ \UR \\$^{4}$ \UoM \\$^{5}$ \IFT \\$^{6}$ \KASSI \\$^{7}$ \LBL \\$^{8}$ \UCL \\$^{9}$ \GRAPPA \\$^{10}$ \CIERA \\$^{11}$ \UPenn \\$^{12}$ \Rice \\$^{13}$ \LLNL \\$^{14}$ \WVU \\$^{15}$ \WVUGWAC \\$^{16}$ \ICE \\$^{17}$ \Nottingham \\$^{18}$ \UCI \\$^{19}$ \CMUCosmo \\$^{20}$ \Queensland \\$^{21}$ \BU \\$^{22}$ \CPPM \\$^{23}$ \SussexAstronomy \\$^{24}$ \ANLHEP \\$^{25}$ \UMich \\$^{26}$ \LUPM \\$^{27}$ \IBS \\$^{28}$ \SMU \\$^{29}$ \EPFL \\$^{30}$ \Brown \\$^{31}$ \BenGurion \\$^{32}$ \INFNFE \\$^{33}$ \Princeton \\$^{34}$ \Tamu \\$^{35}$ \MIT \\$^{36}$ \YorkU \\$^{37}$ \PI \\$^{38}$ \CITA \\$^{39}$ \Siena \\$^{40}$ \IAP \\$^{41}$ \Bohr \\$^{42}$ \Bicocca \\$^{43}$ \UGTO \\$^{44}$ \BNL \\$^{45}$ \WCA \\$^{46}$ \UWaterloo \\$^{47}$ \SoCal \\$^{48}$ \SimonFraser \\$^{49}$ \Stanford \\$^{50}$ \KIPAC \\$^{51}$ \KICP \\$^{52}$ \UChicago \\$^{53}$ \JPL \\$^{54}$ \Caltech \\$^{55}$ \Sejong \\$^{56}$ \Birm \\$^{57}$ \ParisSud \\$^{58}$ \UCLA \\$^{59}$ \NAOC \\$^{60}$ \Port 

\end{document}